\documentclass[twocolumn,showpacs,10pt,prb,floatfix,amssymb]{revtex4-1}
\usepackage{graphicx}

\begin{document}

\newcommand{\nx}{\textrm}

\title{High-order cumulants in the counting statistics of asymmetric quantum dots}

\author{Christian Fricke}
\author{Frank Hohls}
\author{Nandhavel Sethubalasubramanian}
\author{Lukas Fricke}
\author{Rolf J. Haug}
\affiliation{Institut f\"ur Festk\"orperphysik, Leibniz Universit\"at Hannover, 30167 Hannover, Germany.}

\date{\today}
\begin{abstract}
Measurements of single electron tunneling through a quantum dot using a quantum point contact as charge detector have been performed for very long time traces with very large event counts. This large statistical basis is used for a detailed examination of the counting statistics for varying symmetry of the quantum dot system. From the measured statistics we extract high order cumulants describing the distribution. Oscillations of the high order cumulants are observed when varying the symmetry. We compare this behavior to the observed oscillation in time dependence and show that the variation of both system variables lead to the same kind of oscillating response.
\end{abstract}

\maketitle

Current fluctuations in mesoscopic systems allow to obtain information on transport that is not accessible from the average current alone\cite{BlanterButtiker}. Measurements of the shot noise as a first step beyond the mean have been deployed successfully to examine correlations in the electron transport through a metallic island\cite{Birk1995} and through semiconducting  quantum dot systems\cite{nauen2002, nauen2004}, but the extraction of higher moments directly from the current fluctuations is a demanding task \cite{Reulet2003} and has not yet been achieved for transport through quantum dots. In contrast higher moments are naturally accessible in the context of full counting statistics (FCS) \cite{Levitov1996, Bagrets2003}. With the use of a quantum point contact (QPC) as a non-invasive charge detector with sufficiently fast time response, FCS became experimentally feasible in quantum dot physics \cite{Gustavsson2006,Fujisawa2006,Fricke2007,Gustavsson2007}. In a measurement of the 4th and 5th cumulant complex behavior as a function of asymmetry and integration time with local minima was found \cite{Gustavsson2007}. A significantly improved experimental technique has made the extraction of very high-order cumulants possible, revealing an oscillating behavior as function of integration time that strongly increases with the order of the moment\cite{Flindt2009,Fricke2010}.  Such oscillations are also predicted theoretically in quantum optics \cite{ Dodonov1994} and particle physics\cite{Dremin1994}.The theoretical treatment has shown that these oscillations are a universally expected phenomenon for most physical systems and that in quantum dots they should show up prominently as function of the barrier asymmetry\cite{Flindt2008, Flindt2009}.

In this letter we present the experimental verification of the predicted oscillations in the higher cumulants as function of barrier asymmetry in a quantum dot system. To this aim we employ real time single electron counting with a high bandwidth detector. This enables us to measure the full counting statistics for electron transport through a quantum dot on the $\mu$s-timescale, allowing us to gather a unprecedented large amount of counting events. With this large statistical basis we can examine the functional dependence of high order cumulants, which describe the statistical properties of the system, on the tunneling barrier asymmetry and also on the counting time. The observed oscillations as function of both parameters show large similarities that strongly support the universal nature of these oscillations.

Our device is based on a GaAs/AlGaAs heterostructure containing a two-dimensional electron system (2DES) 34 nm below the surface. The electron density is $\rho = 4.6 \cdot 10^{15} \hspace{1.2mm}\mathrm{m}^{-2}$, and the mobility is $\mu = 64 \hspace{1.2mm} \mathrm{m^2/V s}$. We have used an atomic force microscope (AFM) to define the quantum dot (QD) and the quantum point contact (QPC) structure by local anodic oxidation (LAO) of
the surface \cite{held98,keyser2000}; the 2DES below the oxidized surface is depleted and so insulating lines are defined.

\begin{figure}[t!]
\includegraphics[width=\linewidth]{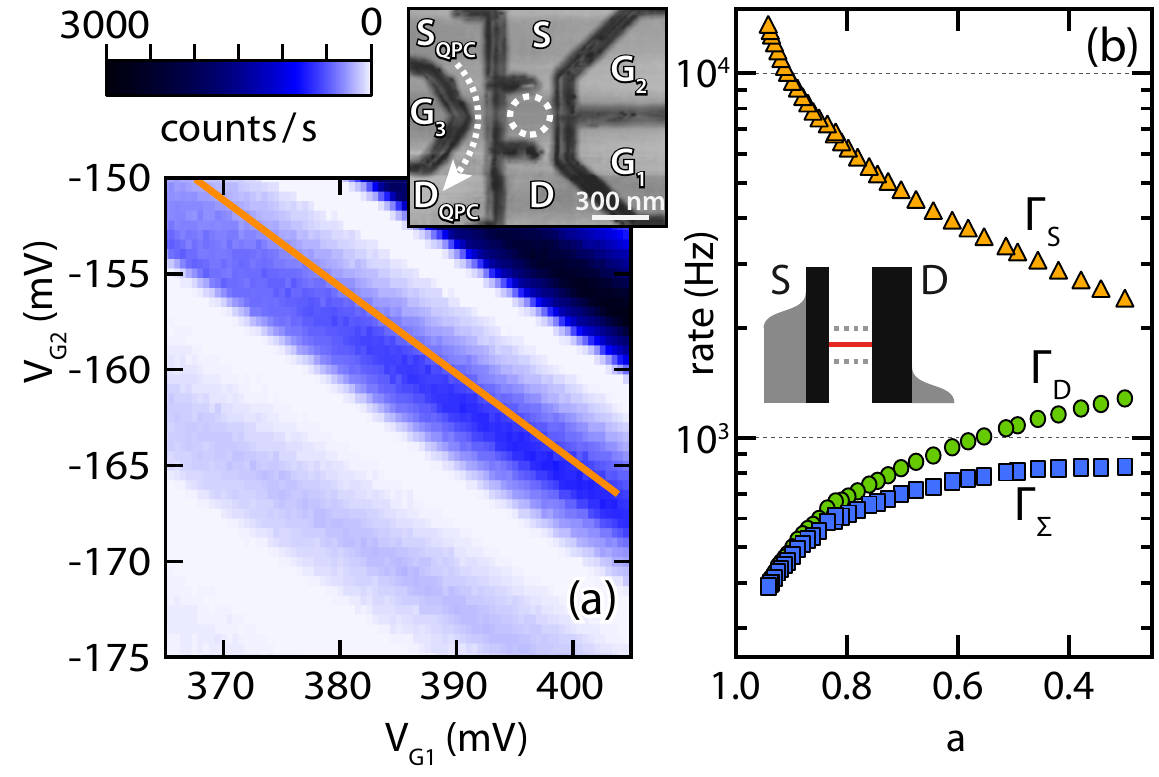}
\caption{\label{fig01}(a) Average current through the quantum dot versus gate voltage at the in-plane gates G1 and G2. Current is measured counting electron by electron. A bias of 1.3 mV is applied between source and drain contact of the quantum dot. Inset: AFM image of the structure. (b) Measured tunneling rates $\Gamma_S$ (source), $\Gamma_D$ (drain) and the electron transport rate $\Gamma_\Sigma$ with respect to the asymmetry parameter $a$. The asymmetry parameter is modified by changing the quantum dot gate configuration along the line shown in (a).}
\end{figure}

An AFM image of our device is presented in Fig.~\ref{fig01}. The dark lines depict the insulating oxide barriers written by the AFM. The QPC (left area) is separated from the QD structure (right area) by an insulating line. The QPC can be electrically tuned using the in-plane gate ${G3}$. The current through the QPC is amplified by a current amplifier and detected in a time-resolved manner with a sampling rate of 500 kHz. To achieve a high bandwidth charge detection a custom low capacitance wiring is used to reduce technical noise caused by the input voltage noise of the amplifier. The bandwidth of the charge detection is only limited by the current amplifier bandwidth and slightly exceeds 100 kHz. The QPC bias is chosen sufficiently small to avoid back-action on the QD \cite{khrapai2006, Gustavsson2007b}.  The QD is coupled to source and drain electrodes via two tunneling barriers which can be separately
controlled with gate voltages $V_{G1}$ and $V_{G2}$. These gates are also used to tune the number of electrons on the QD.  The sample is placed in a He$^3$ refrigerator reaching temperatures down to 400 mK.

Setting the device to a situation where electrons can enter and leave the quantum dot results in a fluctuating charge on the quantum dot. The electron number on the dot is sequentially changing from N to N+1 and from N+1 to N electrons. This leads to a corresponding change of the potential at the QPC which has a working point on the edge of the first conductance step. Thus the QPC acts as a sensitive charge detector as each change in the potential due to a changing electron number on the dot leads to a current change at the QPC.

The average current versus $V_{G1}$ and $V_{G2}$ as measured by single electron counting is shown in Fig.~\ref{fig01}a. A bias of 1.3 mV is applied to the quantum dot. The white areas without electron tunneling events correspond to Coulomb blockade with fixed electron number on the quantum dot. In between wide stripes with a significant number of tunneling events are observed due to the applied QD bias voltage. The charging energy of the quantum dot is 1.6 meV, the single particle levelspacing is 0.27 meV. The nonzero electron temperature leads to an ambiguous tunneling direction at the edges of these stripes where the Fermi-level of a lead is in resonance with the dot level.

To analyze only unidirectional transport, the system is tuned to a situation where the transport state of the dot is sufficiently far from the source and drain Fermi energies, thus allowing no back-tunneling from the dot to the source contact or from drain to the dot (compare inset in Fig.~\ref{fig01}b). The symmetry of the tunneling rates are studied using both quantum dot gates to keep the transport in the unidirectional regime and concurrently to change the asymmetry of the tunneling barriers. The line drawn in Fig.~\ref{fig01}a indicates the gate voltages used in the following. The resulting tunneling rates are displayed in Fig.~\ref{fig01}b with respect to the asymmetry factor $a = (\Gamma_S - \Gamma_D)/(\Gamma_S + \Gamma_D)$ with $\Gamma_S$ and $\Gamma_D$ the tunneling rates between source resp.\ drain and the quantum dot. For smaller $V_{G1}$ and more positive $V_{G2}$ (upper left corner of Fig.~\ref{fig01}a) the system is rather asymmetric with $\Gamma_S\approx $ 20 kHz and  $\Gamma_D\approx $ 0.2 kHz. Following the line $\Gamma_S$ drops down to 2.3 kHz and $\Gamma_D$ increases to 1.2 kHz. Thus the asymmetry factor $a$ is varied from 0.98 to 0.3. The  total rate of electron transfer $\Gamma_\Sigma = \Gamma_S\cdot\Gamma_D/(\Gamma_S + \Gamma_D)$ reaches its highest value $\Gamma_\Sigma=0.9$ kHz at $a=0.3$ and drops with increasing asymmetry $a$.

The detection of individual tunneling events allows us to extract the full counting statistics of electron transport through the quantum dot. Before going into this analysis it is useful to introduce a dimensionless time $\Gamma_S\cdot t$ which is equivalent to setting $\Gamma_S$ equal to unity. This is necessary to separate time and asymmetry dependence as according to Ref.\cite{Flindt2009} time dependent oscillations are universal on the timescale $\Gamma_S\cdot t$. Then the  system is solely governed by the asymmetry parameter $a$. Thus the statistical properties are also completely determined by this single system parameter $a$. Of course we have to note that in our experiment another rate exists, namely the detector bandwidth $\Gamma_\mathrm{det}$. However, if we fix this rate relatively to $\Gamma_S$ we again have a system whose statistical properties are solely varied by the asymmetry. We realize this fixation of the detector bandwidth by a post-selection of detected events with an artificial detector bandwidth of $\Gamma_\mathrm{det} = 4 \cdot \Gamma_S$.

\begin{figure}[t]
\includegraphics[width=\linewidth]{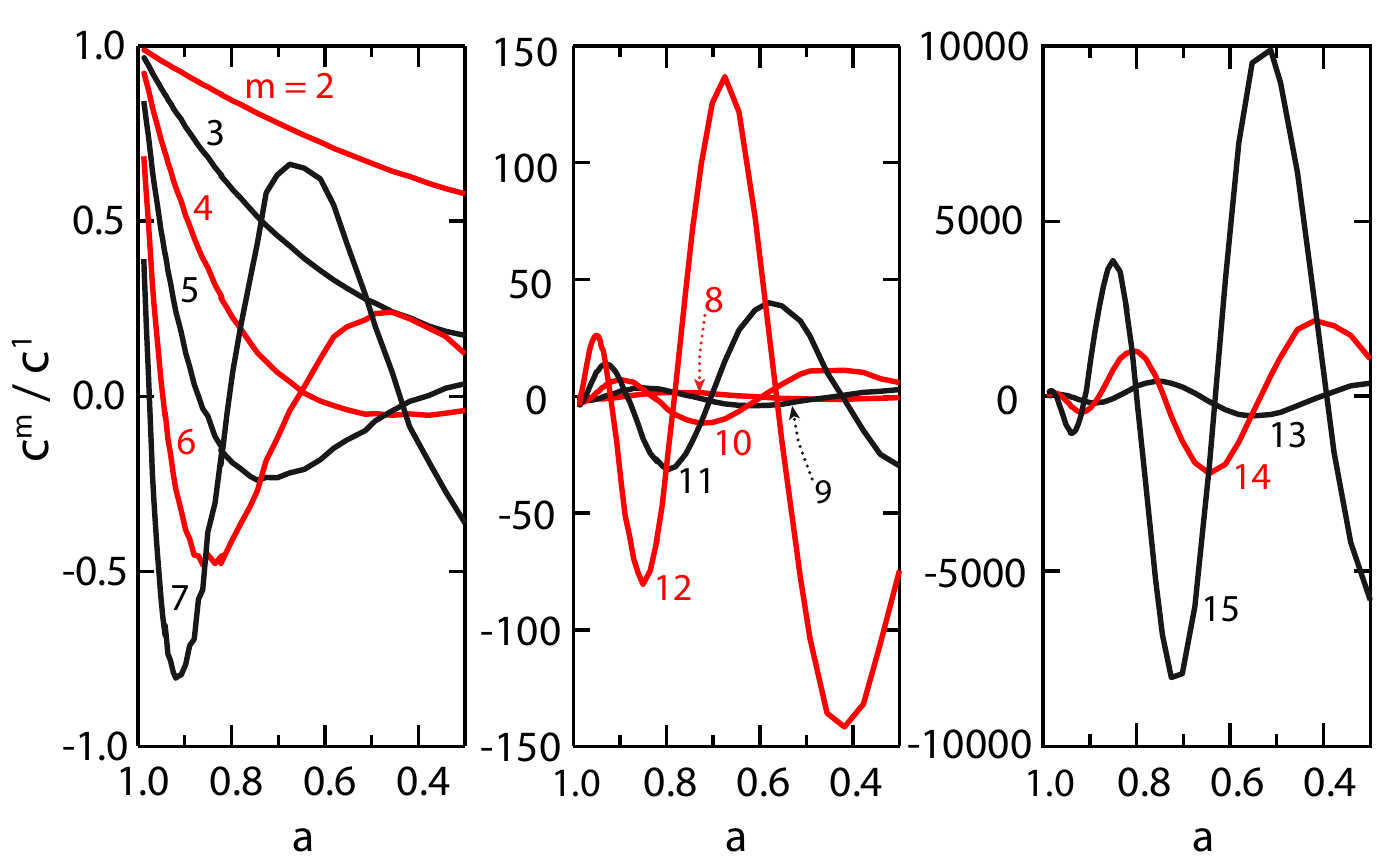}
\caption{\label{fig02} Cumulants of order $m=2\ldots 15$ as function of asymmetry factor $a$. While the low order cumulants show a monotonic behavior the cumulants with order $m > 4$ reveal distinct oscillations with factorial increase in amplitude (up to 10,000 for the 15th cumulant). The cumulants are determined for a constant dimensionless counting time interval $\Gamma_S\cdot t_0$ = 3.2.}
\end{figure}

We will now examine the counting statistics of the quantum dot. This can be done either by direct analysis of the whole probability distribution of the number of transferred electrons $n$ in a certain time interval or more common by the cumulants $c_m$ of this distribution. The first cumulant  is the mean of the number of transferred electrons, $c_1=\left< n \right>$, the second, $c_2$, is the variance, and the third is the skewness. Cumulants of higher order $m$ are very sensitive to the details of the transport process. Recently it was found experimentally that the high-order cumulants of quantum dot transport oscillate as function of the counting time interval.~\cite{Flindt2009} In the same paper it is stated that such oscillations are expected universally for nearly any system when changing a relevant parameter, e.g. the symmetry of the tunneling rates in a quantum dot system~\cite{Flindt2009}. For our experimental study of this predicted symmetry dependence of high-order cumulants  we varied the asymmetry parameter $a$ and measured for each $a$ about 1 million tunneling events. This statistical sample enables us to determine cumulants up to the 15th order with high precision.

In Fig.~\ref{fig02} the normalized cumulants $c^m/c^1$ are shown as function of the asymmetry factor $a$ for $m = 2\ldots 15$. For high asymmetry $(a\rightarrow1)$ all nomalized cumulants start from a value of 1 due to the Poissonian limit of a single dominant barrier. The first cumulants then drop monotonically towards the symmetric limit. For $m \geq 4$ a more complicated behavior can be observed. Beginning with the fourth order cumulant a minimum starts to develop, shifting to higher asymmetry and increasing in amplitude with rising order. For the 6th order an additional maximum appears that also gets more prominent with rising order of the cumulants. This trend continues for even higher cumulants, the number of oscillation increases and also the amplitude rises dramatically. For the 12th order values up to 145 are reached and the amplitude of the 15th cumulant already exceeds 10000. This factorial increase was shown already for oscillations as function of time~\cite{Flindt2009} and appears here again, showing the universal nature of the oscillations.

\begin{figure}[t]
\includegraphics[width=\linewidth]{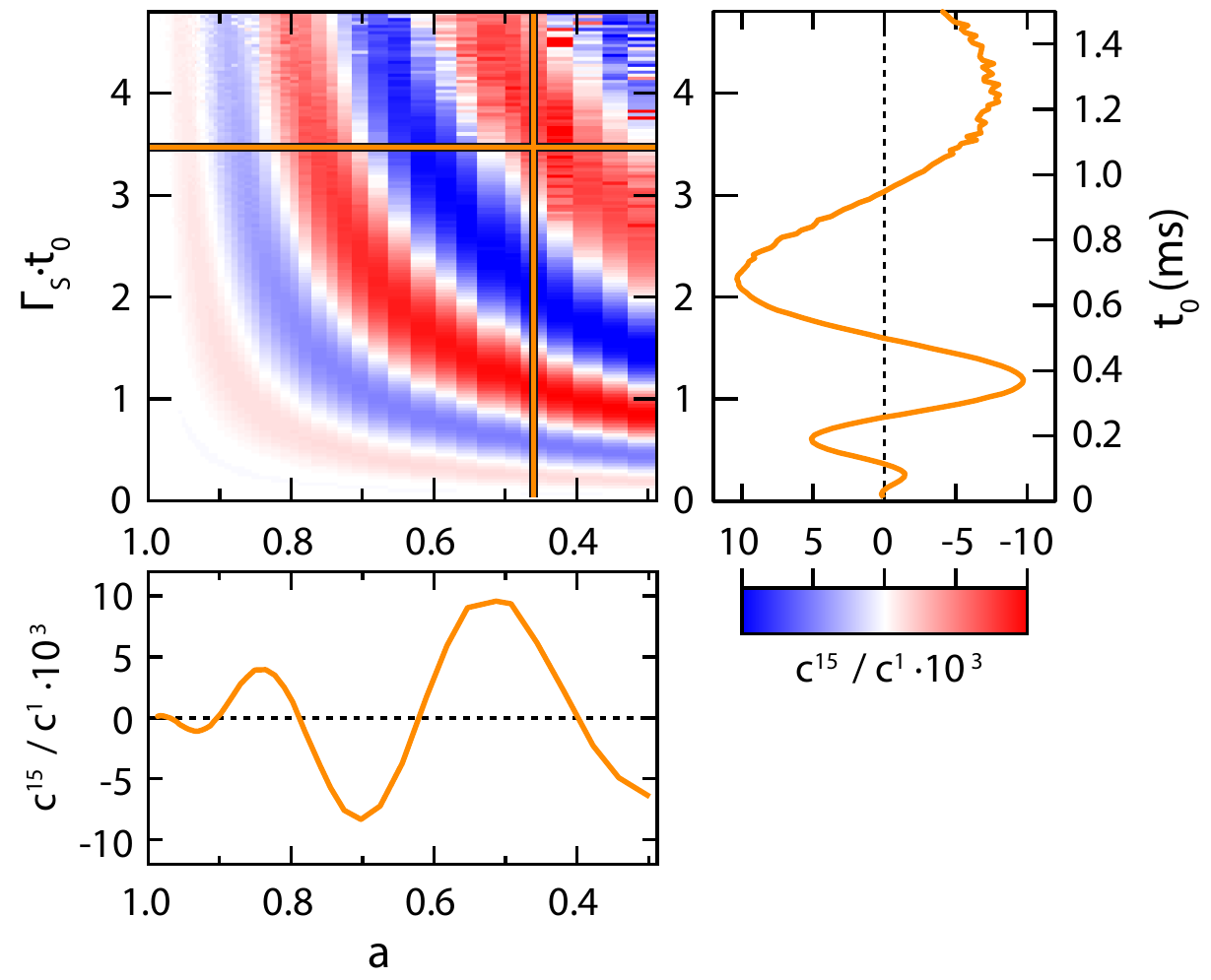}
\caption{\label{fig03} Time and symmetry dependence for the 15th cumulant. Oscillations can be observed for varying time base $\Gamma_S\cdot t_0$ and symmetry $a$. Two profile lines are shown displaying the substantial similarity of both types of oscillations.}
\end{figure}

The asymmetry dependence shown in Fig.~\ref{fig02} was evaluated for a certain choice of the dimensionless length of the counting interval. We can also determine the dependence on the dimensionless counting interval for each value of the asymmetry $a$, thereby mapping out the cumulants as function of both parameters. The result obtained for the 15th cumulant $c_{15}$ is shown in Fig.~\ref{fig03}a. The figure nicely reveals the similarity in the dependencies on asymmetry and on dimensionless time -- the general features are nearly symmetric with respect to the diagonal. For a further illustration of these similarities we plot two profiles in Fig.~\ref{fig03}b and \ref{fig03}c, taken at fixed asymmetry resp.\ fixed normalized time. Both curves, $c_{15}$ as function of time $\Gamma_S\cdot t_0$ resp.\ asymmetry $a$, show oscillations of very similar shape and amplitude. This correspondence between the different parameters nicely demonstrates that the occurrence and the strength of these oscillations are of universal nature~\cite{Flindt2009}. It is also interesting to examine the behavior at long times or towards the case of symmetric rates, $a=0$. For long times the dependence on $\Gamma_S\cdot t_0$ becomes weak and in the long time limit the cumulants will become sole function of the asymmetry. Towards $a=0$ the dependence on asymmetry flattens out and the cumulants depend mainly on time. This transitions again reveal the similarities in the dependence on these different parameters and support the underlying universal nature of the oscillatory behavior.

In conclusion, we have realized a high-bandwidth detection of single electron transport in a quantum dot with a sufficiently large event number to examine the symmetry dependence of the cumulants of the counting probability distribution up to the 15th order. We have verified the theoretically predicted oscillation of the high order cumulants as function of the quantum dot barrier asymmetry. Furthermore we have performed a 2d-parameter study of the cumulants as function of both asymmetry and time. Our experimental results nicely reveal the similarity of the dependence on these different parameters, thereby demonstrating the underlying universal nature of the cumulant oscillations.

We thank C. Flindt for many fruitful discussions. The work was supported by the Federal Ministry of Education and Research of Germany via nanoQUIT and the German Excellence Initiative via QUEST.

\newpage
\end{document}